\documentstyle[12pt,aps]{revtex}

\draft

\draft

\begin{document}


\title{{\small To be published in {\it Journal of Physics A:  Mathematical and General} (2002)} \\
\vskip 0.5cm On the generalized entropy pseudoadditivity for complex systems}


\author{Qiuping A. Wang, Laurent Nivanen, Alain Le M\'ehaut\'e}
\address{Institut Sup\'{e}rieur des Mat\'{e}riaux du Mans, 44, Av.
Bartholdi, 72000 Le Mans, France}
\author{and Michel Pezeril,}
\address{Laboratoire de Physique de l'\'etat Condens\'e,
Universit\'e du Maine,\\
72000 Le Mans, France}

\maketitle

\begin{abstract}
We show that Abe's general pseudoadditivity for entropy prescribed by thermal
equilibrium in nonextensive systems holds not only for entropy, but also for
energy. The application of this general pseudoadditivity to Tsallis entropy
tells us that the factorization of the probability of a composite system into
product of the probabilities of the subsystems is just a consequence of the
existence of thermal equilibrium and not due to the independence of the
subsystems.
\end{abstract}

\vskip 0.5cm


\newpage

Suppose an isolated system composed of two subsystems $1$ and $2$. One of the
basic assumptions of thermodynamics is the existence of an equilibrium state in
the compound system at which $T_1=T_2$, where $T$ is the absolute temperature.
This so called zeroth law is obeyed by Boltzmann statistical mechanics having
additive entropy and energy. For nonextensive statistical mechanics
(NSM)\cite{Tsal99}, the validity of this law is a more subtle
affair\cite{Tsal99,Guer96,Abe99,Abe01,Mart00,Tora01} which depends on the
relationships $aS_{12}=f_{12}(aS_1,aS_2)$ and $bE_{12}=g_{12}(bE_1,bE_2)$,
where $S$ is entropy and $E$ the internal energy. The numbers in the indexes of
the functions indicate the dependence of the latter. $a$ and $b$ are constants
used to make each variable dimensionless in the above equalities. From now on,
we let $a=b=1$ so that $S$ and $E$ become dimensionless.

The zeroth law has been established, in a approximate
way\cite{Abe99,Abe01,Mart00,Tora01}, for NSM with Tsallis entropy and additive
energy $E_{12}=E_1+E_2$. Recently, Abe\cite{Abe01a} furthermore find a general
pseudoadditivity of entropy required by the existence of thermal equilibrium
for additive energy :

\begin{equation}                                    \label{1}
H(S_{12})=H(S_1)+H(S_2)+\lambda_S H(S_1)H(S_2),
\end{equation}
where $H$ is certain differentiable function satisfying $H(0)=0$
and $\lambda_S$ a constant depending on the nature of $S$. For
Tsallis entropy\cite{Tsal99}, $H$ can be proved to be the identity
function. Eq.(\ref{1}) is very interesting because it can be
considered as a general criterion of pertinent nonextensive
entropies for equilibrium systems and may help to understand
Tsallis nonextensive statistical mechanics and to find other
nonextensive thermostatistics obeying the zeroth law.

Nonextensive entropy is a consequence of long range correlations or complex
(fractal or chaotic) space-time. When interactions are no more limited between
or on the walls of the containers of subsystems, energy may be nonadditive.
This issue has been widely discussed and energy nonadditivity was clearly shown
for some cases with long range
interactions\cite{Tsal95,Celia98,Tora01a,Jund95,Cann96,Samp97,Grig96,Lato02,Ruff95}.
So, inspired by Abe's work, we naturally ask the following question : what is
the kind of energy nonadditivity that satisfies the requirement of existence of
thermal equilibrium? In this letter, along the line of Abe, we show that the
above {\it equilibrium pseudoadditivity} holds not only for entropy, but also
for internal energy in more general cases where energy-type thermodynamic
variables are not extensive.

To establish the zeroth law, we make a small variation of the
total entropy $dS_{12}$ given by
\begin{equation}                                    \label{2}
dS_{12}=\frac{\partial f_{12}}{\partial S_1}dS_1+ \frac{\partial
f_{12}}{\partial S_2}dS_2,
\end{equation}
and a variation of total energy $E_{12}$ given by
\begin{equation}                                    \label{3}
dE_{12}=\frac{\partial g_{12}}{\partial E_1}dE_1+ \frac{\partial
g_{12}}{\partial E_2}dE_2.
\end{equation}
At thermal equilibrium, $dS_{12}=0$ should hold and $dE_{12}=0$ results from
energy conservation, leading to
\begin{equation}                                    \label{4}
\frac{\partial f_{12}}{\partial S_1}\frac{\partial S_1}{\partial
E_1}dE_1 = -\frac{\partial f_{12}}{\partial S_2}\frac{\partial
S_2}{\partial E_2}dE_2 ,
\end{equation}
and
\begin{equation}                                    \label{5}
\frac{\partial g_{12}}{\partial E_1}dE_1 = -\frac{\partial
g_{12}}{\partial E_2}dE_2.
\end{equation}
So we obtain
\begin{equation}                                    \label{6}
\frac{\frac{\partial f_{12}}{\partial S_1}}{\frac{\partial
g_{12}}{\partial E_1}}\frac{\partial S_1}{\partial E_1} =
\frac{\frac{\partial f_{12}}{\partial S_2}}{\frac{\partial
g_{12}}{\partial E_2}}\frac{\partial S_2}{\partial E_2}.
\end{equation}
Equilibrium means that this above equation yields the zeroth law
which is in general given by following equality :
\begin{equation}                                    \label{7}
F_1=F_2
\end{equation}
where $F_1$ and $F_2$ are the same functions depending on
subsystem-$1$ and $-2$, respectively. This constraint from the
zeroth law implies following factorizations of the derivatives in
Eq.(\ref{6}) :
\begin{equation}                                    \label{8}
\frac{\partial f_{12}}{\partial S_1}=\phi_{12}\omega_1\nu_2,
\end{equation}

\begin{equation}                                    \label{9}
\frac{\partial f_{12}}{\partial S_2}=\phi_{12}\nu_1 \omega_2,
\end{equation}

\begin{equation}                                    \label{10}
\frac{\partial g_{12}}{\partial E_1}=\theta_{12}\mu_1\upsilon_2,
\end{equation}
and
\begin{equation}                                    \label{11}
\frac{\partial g_{12}}{\partial E_2}=\theta_{12}\upsilon_1\mu_2.
\end{equation}
where $\phi$, $\omega$, $\nu$, $\theta$, $\upsilon$ and $\mu$ are certain
functions of the subsystems indicated by the indexes. From Eq.(\ref{6}) we can
write
\begin{equation}                                    \label{12}
\xi_1\frac{\partial S_1}{\partial E_1} =\xi_2\frac{\partial
S_2}{\partial E_2}.
\end{equation}
with $\xi_1=\frac{\omega_1\upsilon_1}{\nu_1\mu_1}$ and
$\xi_2=\frac{\omega_2\upsilon_2}{\nu_2\mu_2}$. This shows that
Eq.(\ref{8}) to (\ref{11}) are really the most general forms of
the derivatives of $f$ and $g$ satisfying Eq.(\ref{7}).

We find that all the calculations of Abe\cite{Abe01a} hold for $f$
as well as for $g$. So we can replace $S$ by $E$ in Eq.(\ref{1}).
For a system containing $N$ subsystems in equilibrium, we have
\begin{equation}                                    \label{1c}
\ln[1+\lambda_x H_x(x_{12..N})]=\sum_{i=1}^N\ln[1+\lambda_x
H_x(x_i)].
\end{equation}
where $x$ can be entropy $S$ or energy $E$. In general, $H_S$ and
$\lambda_S$ are different from $H_E$ and $\lambda_E$,
respectively.

We mention here as example two nonextensive cases with Tsallis entropy
($S=-\frac{1-\texttt{Tr}\rho^q}{1-q}$, $\rho$ is density operator) where the
zeroth law is claimed to be verified.
\begin{enumerate}
\item
In the Tsallis nonextensive statistics with escort probability,
$H_S$ and $H_E$ are the identity function, $\lambda_S=(1-q)$ and
$\lambda_E=0$\cite{Tsal99,Abe99,Abe01,Mart00,Tora01} (i.e. $E$ is
extensive and $S_{12}=S_1+S_2+(1-q)S_1S_2$).

\item Another possible case is with Tsallis entropy combined with a so called
incomplete normalization\cite{Wang01,Wang01c,Wang01b} where $H_x$ is the
identity function and $\lambda_x=q-1$ for both entropy and energy (i.e.
$x_{12}=x_1+x_2+(q-1)x_1x_2$ or, according to Eq.(\ref{1c}),
$x_{12..N}=\frac{[\prod_{i=1}^N(1+\lambda_x x_i)-1]}{\lambda_x}$ with $N$
subsystems. In this case, the zeroth law can hold without approximation, making
it possible to establish an exact nonextensive thermodynamics.
\end{enumerate}

An example of systems satisfying pseudoadditivity of entropy is given by
Abe\cite{Abe01a} with $H_x(x)=\sqrt{x}$ for black hole entropy proportional to
its horizon area. This discussion should hold for energy as well according to
the first law of thermodynamics for black hole\cite{Hayw98} if electromagnetic
work is absent. So it would indeed be interesting to study black hole within a
generalized thermodynamics with nonadditive entropy and energy as well in view
of the difficulty with Boltzmann statistics due to the presence of thermal
(infrared) divergence\cite{Frol98}. Another example can be given with the
long-range ferromagnetic spin model of which the internal energy is given by
$E(N,T)=c(T)N\frac{N^{1-\alpha/d}-1}{1-\alpha/d}$\cite{Tsal95,Tora01a,Cann96}
where $N$ is the number of particles in the model supposed additive, $d$ the
dimension of space, $\alpha$ the exponent in the factor $1/r^\alpha$ of the
long range potential\cite{Tsal95,Celia98,Tora01a,Jund95,Cann96,Samp97,Grig96}
and c(T) certain function of temperature $T$. When $N\rightarrow \infty$ and
$\alpha
>d$, $E(N,T)=c(T)N$ is additive. On the other hand, when $\alpha=d$,
$E(N,T)=c(T)N\ln N$. If we put, e.g., $H_E(E)=\{exp[e^{E/c(T)N}]-e \}/e$ and
$\lambda_E=1$, then $E$ satisfies Eq.(\ref{1}) or Eq.(\ref{1c}). When
$0<\alpha<d$, $E(N,T)=c(T)N^{2-\alpha/d}$, we can choose
$H_E(E)=[\frac{E}{c(T)}]^{1/(2-\alpha/d)}$ and $\lambda_E=0$ for energy to
satisfy Eq.(\ref{1}) or Eq.(\ref{1c}). There exist other choices, e.g.
$H_E(E)=\exp \{[\frac{E}{c(T)}]^{1/(2-\alpha/d)}\}-1$ and $\lambda_E=1$.

As a matter of fact, it seems that, for a given explicit relation between
energy or entropy and number of subsystems or volume supposed additive (i.e.
$V_{12}=V_1+V_2$ and $N_{12}=N_1+N_2$), the finding of a function $H$
satisfying Eq.(\ref{1}) is a trivial affair. The most essential contribution of
Abe's work, in our opinion, is that Eq.(\ref{1}) finally makes it clear that
the factorization of the compound probability of a composite system into
product of the probabilities of the subsystems is a consequence of the
existence of thermodynamic equilibrium {\it if Tsallis entropy applies},
because $H$ is identity function here. It is straightforward to show that :
$S_{12}=S_1+S_2+(1-q)S_1S_2 =-\frac{1-\texttt{Tr}\rho_1^q}{1-q}
-\frac{1-\texttt{Tr}\rho_2^q}{1-q}+ (1-q) \frac{1-\texttt{Tr}\rho_1^q}{1-q}
\frac{1-\texttt{Tr}\rho_2^q}{1-q} =-\frac{1-\texttt{Tr}(\rho_1\rho_2)^q}{1-q}
=-\frac{1-\texttt{Tr}\rho_{12}^q}{1-q}$, which means $\rho_{12}=\rho_1\rho_2$
[or that $p_{ij}^q(1,2)=p_i^q(1)p_j^q(2)$ implies $p_{ij}(1,2)=p_i(1)p_j(2)$
where $p_i(1)$ or $p_j(2)$ is the probability for subsystem-1 or -2 to be at
state $i$ or $j$ and $p_{ij}(1,2)$ is the probability for the composite system
to be at the product state $ij$]. So {\it the product probability is rather due
to thermal equilibrium instead of independence of noninteracting or weakly
interacting subsystems as claimed or insinuated in most of the current
publications}. On this basis, we can finally free ourselves from the paradox of
addressing {\it noninteracting independent systems} with {\it nonadditive
entropy due to long range interactions}. Additive energy based on independent
systems with only short range interactions does not conform with the spirit of
nonextensive statistical mechanics.

The law of product probability must be respected for any thermodynamic system
in equilibrium. This means that, for NSM with the Tsallis $q$-exponential
distribution, nonextensive energy is needed for {\it exact treatments of
nonextensive interacting systems}. In this sense, all treatments within NSM
with additive energy should be viewed as a kind of {\it extensive
approximation} and should be proceeded with great care because they give
sometimes very different results from the treatment respecting probability
factorization or product state\cite{Wang01b,Lenz01}. This above ``equilibrium
interpretation" of the factorization hypothesis of compound probability may
have important consequences on the applications of NSM to many-body systems.
More detailed discussions on this issue are given in others papers of
ours\cite{Wang02}.

Summing up, we have applied Abe's method of finding general entropy
pseudoadditivity to a more general case where both entropy and energy are
nonextensive. Under thermal equilibrium, the energy of a nonextensive composite
system also obeys Abe-type pseudoadditivity.

We would like to thank Professor Sumiyoshi Abe for critical reading of this
manuscript and important comments. Thanks are also due to Professor S. Ruffo
for valuable discussions.

\end{document}